# Angle-resolved broadband ferromagnetic resonance apparatus enabled through a spring-loaded sample mounting manipulator

Shikun He,[1,2,a] Qing Qin[1], Tiejun Zhou[1] and Christos Panagopoulos[2,b]

[1]*Data Storage Institute, A*STAR (Agency for Science Technology and Research), 2 Fusionopolis Way 08-01 Innovis, Singapore 138634*
[2]*Division of Physics and Applied Physics, School of Physical and Mathematical Sciences, Nanyang Technological University, Singapore 637371*

## Abstract

Broadband ferromagnetic resonance is a useful technique to determine the magnetic anisotropy and study the magnetization dynamics of magnetic thin films. We report a spring-loaded sample loading manipulator for reliable sample mounting and rotation. The manipulator enables maximum signal, enhances system stability and is particularly useful for fully automated in-plane-field angle-resolved measurements. This angle-resolved broadband ferromagnetic resonance apparatus provides a viable method to study anisotropic damping and weak magnetic anisotropies, both vital for fundamental research and applications.

[a] heshikun@gmail.com
[b] christos@ntu.edu.sg

Manipulating magnetization through electrical currents via spin-transfer torque and spin-orbit torque is superior to conventional schemes such as the usage of magnetic fields due to its scalability and energy efficiency[1]. Consequently, emerging applications based on spin torques have attracted considerable attention. In particular, it has been demonstrated that spin-transfer-torque magnetic random access memory has the potential to compete with mainstream memory technologies[2,3]. Magnetic thin films and multilayers with low damping constant, and high anisotropy are essential for reducing the switching current density without scarifying data retention performance.

Broadband ferromagnetic resonance (FMR) is widely used for studying the mechanism of magnetization relaxation in ultrathin films[4,5]. It has been shown that the damping parameter can be determined accurately in a field perpendicular to the plane geometry[6,7], assuming that damping is isotropic. However, both phenomenological theory and recent experiments indicate that the Gilbert damping constant can depend on the orientation of the magnetization vector[8,9]. Crucially, prior to the calculation of anisotropy in damping, it is important to determine the extrinsic contributions to the FMR linewidth such as inhomogeneous broadening and magnetic anisotropy[7,10] therefore, it is desirable to develop an angle-resolved FMR apparatus to measure spectra at various magnetization orientations and frequencies.

Here, we present a fully automated angle-resolved broadband FMR design which meets those requirements. The key component of the apparatus is an in-plane sample mounting manipulator, which incorporates a spring-loaded sample holder and is attached to a 2-axis motorized stage for automation.

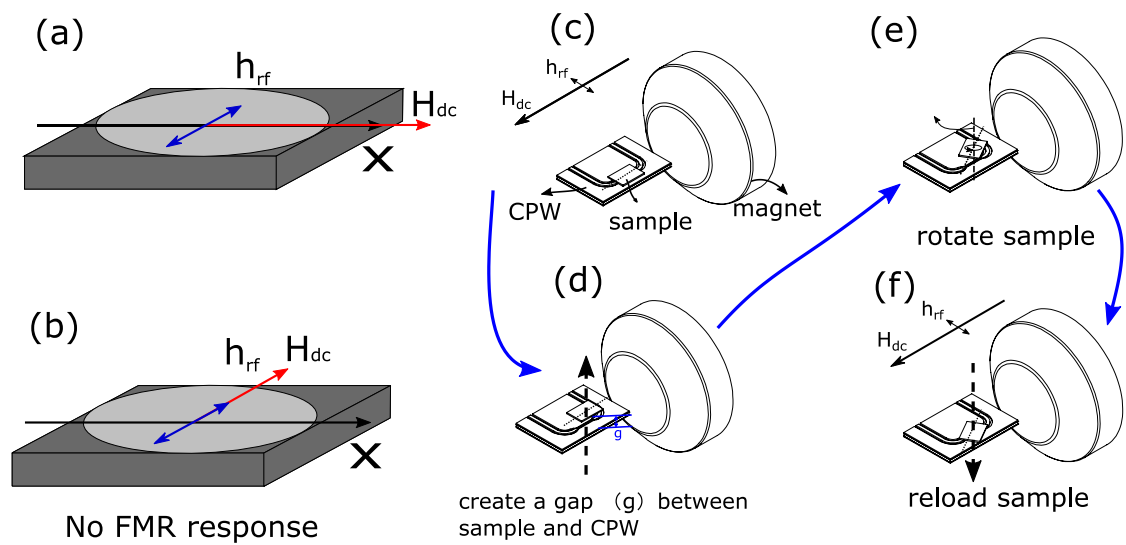

Figure 1: Comparison of two different schemes for achieving field- in-plane angle-resolved FMR measurements. (a), (b) Field rotation scheme. (c)-(f) Sample rotation scheme. First, lift the sample up to create a gap between sample and CPW. Then, rotate the sample to the desired orientation. Finally, reload the sample.

In the broadband FMR technique, the thin film sample is flipped and placed on top of a coplanar waveguide (CPW). The gap between the film and the CPW is minimized to ensure that the microwave field ($h_{rf}$) generated by the RF-current-flow in the central conductor of the CPW is applied effectively on the sample under investigation. Consequently, the microwave field will always be in the sample plane. Figure 1 depicts the two possible schemes to vary the orientation of the magnetization of the sample to enable field-in-plane angle-resolved FMR. The first scheme allows a change in the orientation of the external field ($H_{dc}$) as shown by Figs. 1a and 1b. This method is straightforward to implement by using a vector-magnet (if rotating the magnet base is cumbersome). Because the precession of magnetization is caused by the component of $h_{rf}$ perpendicular to the external field $H_{dc}$, there will be no FMR signal for $H_{dc}||h_{rf}$ (Fig. 1b). Furthermore, there will be an unavoidable and significant reduction of signal when $H_{dc}$ is tilted away from the x-axis. Therefore, such configuration would hinder the investigation of ultra-thin films. In the second scheme one rotates the sample, keeping the waveguide and the direction of $H_{dc}$ fixed. This scheme is preferred from a sensitivity perspective because $H_{dc}$ and $h_{rf}$ remain perpendicular to each other (Figs. 1c and 1f). However, it is necessary to decouple the contact between the sample and the CPW before each sample rotation in order to avoid scratching the sample and the CPW. Hence, one would follow the procedures illustrated in Fig. 1c to 1f to rotate the sample. It is well known that the sample mounting method has a significant impact on the signal strength and noise level of FMR spectrum. Hence, a reliable loading and unloading method is crucial to the performance of this technique. To meet these stringent requirements, we developed an in-plane sample mounting manipulator.

As shown in Fig. 2a, the axis of the in-plane sample mounting manipulator is aligned to the center of the FMR probe head where a CPW is mounted. The FMR probe head (Fig. 2b) is attached to a separate supporting rod, the axis of which is perpendicular to the in-plane sample mounting manipulator. Prior to activating the in-plane manipulator, the orientation of the FMR probe head is set to the default value stored in the configuration data file by the associated rotary positioner. The manipulator is attached to a rotary positioner, which controls the sample orientation. The rotary positioner, in turn, is put into a linear positioner to control the height of the manipulator and hence the location of the sample.

The in-plane rotation procedure is as follows: First, attach the sample (item 11 of Fig. 2d) to the bottom of the sample holder using glue or double-sided tape. Then, mount the sample holder to the manipulator just as mount a screw. Next, the software will follow the procedure listed in Fig. 1c to Fig. 1f to measure FMR spectra at specified in-plane field orientations. To avoid damaging the CPW and the sample, the sample holder height is controlled by a Cu spring (item 5 of Fig. 2d) and two slots (item 3 of Fig. 2d). The usage of the spring minimizes the gap between the sample and the CPW with a controllable force (Fig. 2a), which is essential for the reliability of this manipulator. The tolerance of the travel distance is about

5 mm, which is orders magnitude larger than the control precision of commercially available linear positioners. The difference between the width of the guiding slots and the outer diameter of the set screw is 0.05 mm. Thus, the maximum error of the sample orientation is 0.7 degrees, as estimated by the outer diameter of the housing tube (25 mm, item 4 of Fig. 2d).

The procedure for mounting the sample in the out-of-plane rotation measurement using the probe head only (Fig. 2b) is as follows: First, the in-plane sample mounting manipulator is deactivated and pulled up automatically by the linear positioner. Next, we place the flipped sample at the center of the CPW (Fig. 2c). Third, we press the clip base (item 6), rotate the spring clip (item 5) to a location on top of the sample and release the clip base. The out-of-plane measurement through rotating the FMR probe head is trivial so therefore needn't be discussed.

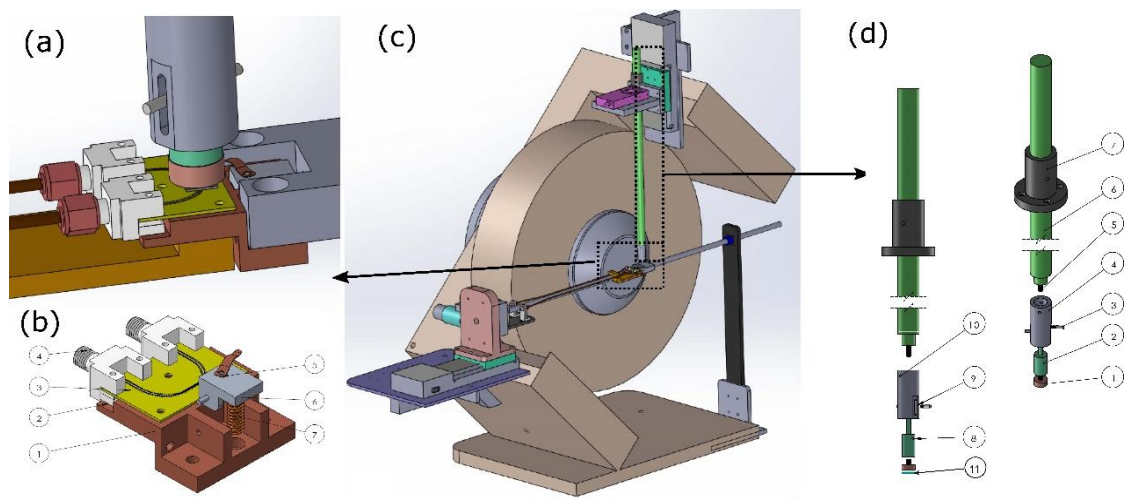

Figure 2: (a) A schematic of the FMR head and sample location during in-plane rotation. (b) The FMR probe head. 1-Copper housing; 2-Endlaunch connector; 3-coplanar wave guide (CPW); 4-2.4mm cable connection port; 5-BeCu spring clip; 6-clip base; 7-Cu spring (c) The angle-resolved FMR system based on an electromagnet. The in-plane sample mounting manipulator is shifted back for clarity. (d) The in-plane sample mounting manipulator. 1-sample holder; 2-spring holder; 3-set screw; 4-housing tube; 5-Cu spring; 6-shaft; 7-adapter to positioner; 8-screw hole for item 3; 9-guiding slot; 10-mounting screw hole; 11-sample.

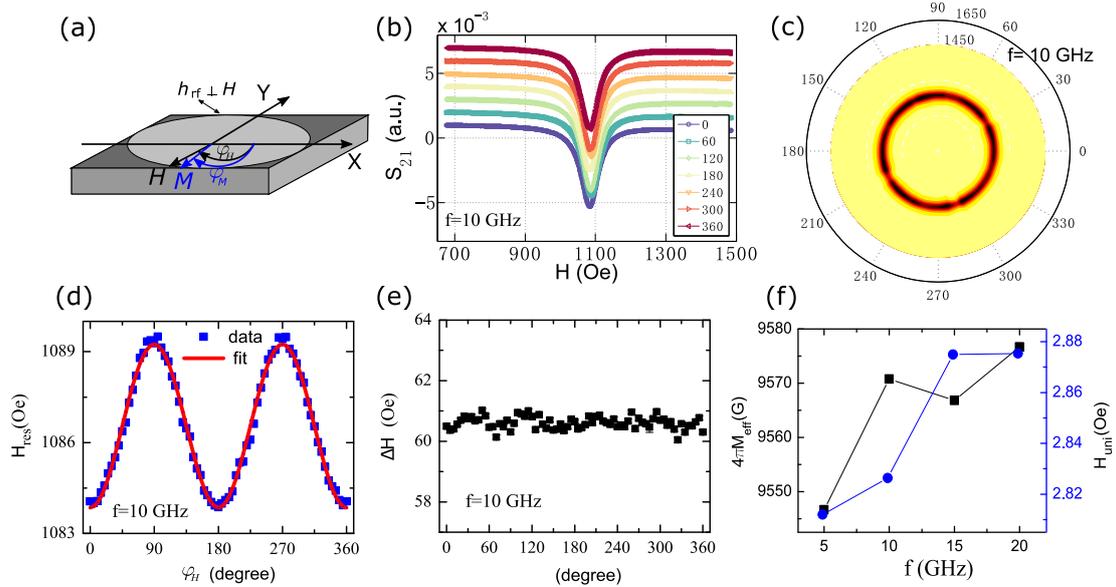

Figure 3: Results of a Si/Py 10nm/Ta 5nm sample. (a) Schematic of the coordinate system used for the in-plane rotation. (b) FMR raw data at various field orientations for f=10 GHz. (c) 2D mapping of the FMR spectra at 10 GHz. (c) The field orientation $\varphi_H$ dependence of the resonance field. (e) The FMR linewidth versus field orientation $\varphi_H$. (f) The effective saturation magnetization and anisotropy field value calculated at various frequencies.

Shown in Fig. 3 is the in-plane FMR data obtained for a 10 nm Py sample prepared by magneto-sputtering. Resonance lineshape is observed for all of the field/magnetization orientations. The nearly isotropic FMR response is shown in Fig. 3c. Solving the Landau–Lifshitz–Gilbert equation in polar coordinates[11,12], we derive the general formula of dynamic susceptibility for the in-plane angle-resolved FMR:

$$\chi = \frac{4\pi M_s}{PQ-(2\pi f/\gamma)^2 + i\frac{\Delta H\cos(\varphi_M-\varphi_H)}{2}(P+Q)}[P+i\frac{\Delta H\cos(\varphi_M-\varphi_H)}{2}] \quad (1)$$

Here, $P=\frac{1}{M}\frac{\partial^2 E}{\partial \theta_M^2}$, $Q=\frac{1}{M\sin^2\theta_M}\frac{\partial^2 E}{\partial \phi_M^2}$ are the second partial derivatives of total free energy density $E$. However, fitting the real and imaginary part of the spectrum to Eq.1 simultaneously requires prior knowledge of $E$, which is inconvenient. Alternatively, to determine the FMR linewidth accurately without prior information on the actual anisotropy energies, we fit the data to a Lorentz lineshape:

$$S_{21} = \frac{A\cdot(\Delta H/2)^2}{[(H-H_{Res})^2+(\Delta H/2)^2]} + \frac{B\cdot\Delta H/2\cdot(H-H_{Res})}{[(H-H_{Res})^2+(\Delta H/2)^2]} + C \quad (2)$$

where $A$ and $B$ are the amplitude of symmetric and antisymmetric lineshape, respectively, $C$ is the background of the spectrum, $H_{Res}$ is the resonance field and $\Delta H$ is the full width at half maximum. We find a weak two-fold symmetry in the magnetization dependence of $H_{res}$ (Fig. 3d) and therefore, we can determine the resonance condition by adding a uniaxial anisotropy term in the total energy and solving the Smit-Beljers equation[11]:

$$2\pi f = \gamma\sqrt{PQ}$$
$$= \sqrt{(H_{res}\cos(\varphi_M-\varphi_H)+4\pi M_{eff}+H_{uni}\cos^2\varphi_M)(H_{res}\cos(\varphi_M-\varphi_H)+H_{uni}\cos 2\varphi_M)} \quad (3)$$

Here, $\varphi_M$ is calculated numerically by searching for the energy minimum of $E$ in a single domain approximation. Notably, $4\pi M_{eff}$ and $H_{uni}$ are material related parameters, and do not change with measurement frequency. As shown in Fig. 3f, the values determined are nearly identical for all the frequencies used. Remarkably, the small anisotropy, which is less than 3 Oe, possibly due to steps in the substrate, can be determined with a standard deviation of less than 5%. The FMR linewidth at a given frequency is also essentially the same for all magnetization orientations, as expected by the isotropic nature of the Py film (Fig. 3e).

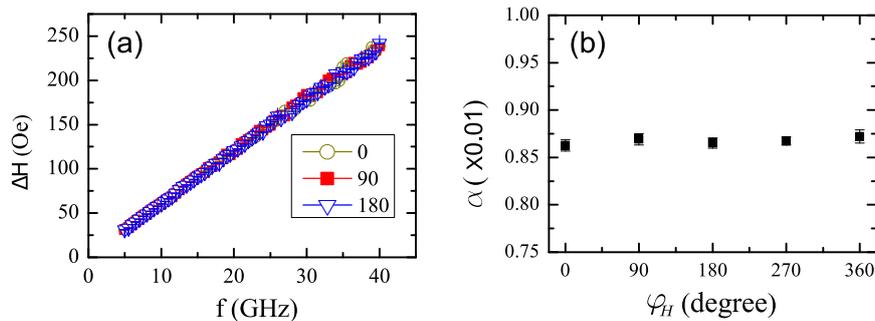

Figure 4: (a) Frequency dependence of FMR linewidth of a Si/Py 10nm/Ta 5nm film. (b) The extracted damping constant for the film at various field orientations.

Figure 4a depicts the frequency dependence of the FMR linewdith. The observed linearity for any field orientation allows the calculation of damping along each field direction $\Delta H = \frac{4\pi}{\gamma}\alpha f + \Delta H_0$. Here, we find the damping constant for Py to be isotropic (Fig. 4b). Notably, besides angle-dependent damping one can use the nonlinear linewidth versus frequency curves, which are typical for ultra-thin films, to extract the extrinsic contributions to magnetization relaxation and the associated angle-dependence.

In summary, we developed a spring-loaded sample loading manipulator for reliable and fast angle-resolved broadband FMR measurements. We demonstrated the suitability of the apparatus for determining the magnetic anisotropy with high precision and measuring the damping constant at all in-plane magnetization orientations. Importantly, such a manipulator can be designed in a variety of configurations to meet the geometrical constraints of a particular system.

**Acknowledgements**

CP acknowledges support from the National Research Foundation (NRF), NRF-Investigatorship (No. NRFNRFI2015-04).